# Axiomatic Incompatibility Between Thermal Equilibrium and Volume-Dependent Entropy in Ideal Gas Theory

A. Paglietti

University of Cagliari, 09123 Cagliari, Italy
E-mail: paglietti@unica.it
Orcid ID: 0000-0002-5294-6012

**Abstract**

A rigorous macroscopic analysis of classical thermodynamics reveals an overlooked contradiction at the foundations of ideal gas theory: the traditional volume-dependent formula for entropy violates the fundamental axiom of thermal equilibrium during reversible isothermal processes. This paper resolves this inconsistency by establishing a strict operational distinction between strictly isothermal and quasi-isothermal processes. We show that in order for the perfect gas to be strictly consistent with the laws of classical thermodynamics, the thermodynamic entropy of a perfect gas must be a function of temperature only. This independence of volume is also confirmed by a new analysis of the adiabatic free expansion of an ideal gas, which is found to be intrinsically reversible. To reconcile these results with standard thermodynamic practice, we introduce the concept of a pseudo-isothermal process—a finite sequence of small adiabatic and isochoric steps—and show that the classical equations for work and heat apply rigorously to these real processes, as opposed to strictly isothermal ones. We conclude by discussing the profound implications of these findings for statistical mechanics, showing that while the corrected thermodynamic entropy is in perfect agreement with Maxwell's kinetic theory, it is essentially different from the Planck–Boltzmann formula. Thus, thermodynamic entropy and statistical microstate-counting are two different physical quantities, and the present analysis provides a simple way to free classical thermodynamics from the paradoxes of microscopic disorder.

## 1. Introduction

Ideal gases are fundamental to thermodynamics; their simplicity offers valuable insights into the classical and statistical approaches to this discipline. However, the traditional approach to ideal gas theory contains some inconsistencies that have so far remained unnoticed or have not been properly addressed. This work reexamines the thermodynamics of ideal gases from a rigorous macroscopic perspective, following the basic premise that no heat transfer can occur between systems in thermal equilibrium at identical temperatures. It will be shown here that the classical equation for the thermodynamic entropy of ideal gases needs to be revised.

In most of the literature on this subject, there is a widespread practice of treating as isothermal those processes in which the system's temperature is not strictly constant but undergoes only small variations around a fixed value. This may be adequate for some applications, but from a strictly thermodynamic standpoint, any temperature variation, even an infinitesimal one, can create a finite temperature gradient with the surrounding environment. In this case, if the system is not adiabatically insulated, spontaneous heat exchange becomes inevitable. A similar situation can never occur in a reversible strictly isothermal process: no heat exchange is possible if the system and its surroundings remain in thermal equilibrium at the same temperature.

To avoid thermodynamic inconsistencies, it is therefore necessary to draw a clear distinction between strictly isothermal processes, on the one hand, and quasi-isothermal processes, on the other. Failure to do so could lead to inaccurate theoretical conclusions, as the present analysis will show.

In this paper, it will be demonstrated that an isothermal, reversible process of an ideal gas is also adiabatic and workless. It will be shown that this implies that its entropy must be volume-independent, since the internal energy of the ideal gas is independent of its volume. However, the accepted formulae for the entropy of an ideal gas in classical and statistical mechanics say otherwise. This conflict calls for a careful reassessment of how entropy is understood and calculated, especially concerning its connection with microscopic disorder.

The remainder of the paper is organized as follows. Section 2 gives some basic definitions and equations that underpin the thermodynamic theory of ideal gases. In Section 3, a clear distinction is drawn between strictly isothermal and quasi-isothermal processes. Section 4 demonstrates that the basic principles of thermodynamics, if strictly applied, forbid an ideal gas to exchange heat in a process that is both strictly isothermal and reversible. This is not the case if a reversible process is quasi-isothermal. Although quasi-isothermal processes are often treated as isothermal, they still involve temperature fluctuations and can therefore lead to heat transfer.

The restrictions on the entropy of an ideal gas that are imposed by the laws of thermodynamics are analyzed in Section 5, and it is demonstrated that the entropy of an ideal gas cannot depend on volume. Two independent proofs are provided. An explicit expression for the entropy of an ideal gas is given in Section 6.

The classical formulae for the work and heat absorbed by an ideal gas in a reversible isothermal process represent some of the most reliable results of classical thermodynamics. The conclusion given in Section 3 that in a reversible isothermal process, an ideal gas can neither do work nor absorb heat might appear to conflict with these results. In Section 7, this conflict is addressed by introducing the concept of a pseudo-isothermal process. This process is not strictly isothermal; however, it allows for temperature variations that can be maintained very close to a constant value. The classical formulae traditionally attributed to reversible isothermal processes are still valid in this more rigorous framework, provided that they are used to refer to a reversible pseudo-isothermal process rather than (incorrectly) to an isothermal process.

Section 8 describes the adiabatic free expansion of an ideal gas and shows that, contrary to popular belief, the process must be reversible. The way in which an expanded gas can be returned reversibly to its original pre-expansion state is described. This finding provides further support for the expression for the entropy for an ideal gas derived in Section 5. A new expression for the free energy of an ideal gas is calculated in Section 9, and its consistency with Maxwell's symmetry relations is discussed.

Finally, Section 10 examines the implications of the present analysis for the kinetic theory of gases and statistical mechanics. It is shown that the macroscopic formula for the entropy of an ideal gas obtained in Section 6 is fully consistent with the entropy formula implied by the kinetic theory of gases. It is interesting to note that entropy was not taken into account by Maxwell when he proposed his kinetic theory.

It is also pointed out in Section 10 that the Planck–Boltzmann equation for entropy, which is widely in use in statistical mechanics, is volume-dependent and hence cannot represent the same entropy that is introduced in macroscopic thermodynamics. This finding reinforces the need for re-evaluation of the meaning and implications of statistical entropy [1]-[3]. It should also help to liberate classical thermodynamics from the unwarranted influence of the notion of microscopic disorder, which is not consistent with a macroscopic approach [4]-[7].

The analysis presented in this paper does not invalidate classical thermodynamics; instead, it corrects the misapplication of some of its basic concepts. Classical theory is also reconciled with Maxwell's kinetic theory, and inconsistencies in the commonly accepted links between thermodynamics and statistical mechanics are identified.

## 2. Thermodynamic Preliminaries

Throughout this paper, we will always refer to homogeneous systems at a uniform temperature. In this case, the differential forms of the first and second laws of thermodynamics can be written, respectively, as

$$\mathrm{d}U = đW + đQ \tag{1}$$

and

$$\mathrm{d}S \geq \frac{đQ}{T} \; . \tag{2}$$

Here, $W$ is the work done *on* the system and $Q$ the heat absorbed *by* the system. The state functions $U$ and $S$ are the internal energy and entropy of the system. The differential symbol đ serves to warn that đ$Q$ and đ$W$ are not exact differentials; this means that their integrals $Q$ and $W$ are dependent on the system's process.

A reversible process is defined as a sequence of equilibrium states (a quasi-static process) that occur in thermal equilibrium with the environment. This process can be reversed without causing any changes to the system or its surroundings, which are both returned to their original states (cf., for example, p. 19 of Ref. [8] or p. 12 of Ref. [9]).

The equality sign in Eq. (2) applies if and only if the process is reversible. Hence, for reversible processes, the relation in Eq. (2) is reduced to

$$\mathrm{d}S = \frac{đQ}{T} \; . \tag{3}$$

This equation is therefore both necessary and sufficient for any infinitesimal step in a process to be reversible.

The above equations are fundamental to classical thermodynamics. The entropy referred to here is thermodynamic entropy. Its variation in the change from one state A to another state B is calculated by integrating Eq. (3) along any reversible process that joins the two states:

$$S_\mathrm{B} - S_\mathrm{A} = \int_\mathrm{A}^\mathrm{B} \frac{đQ}{T}\bigg|_\mathrm{rev} \; . \tag{4}$$

The state of an ideal gas is defined by its volume, $V$, and temperature, $T$. Its internal energy is a function only of temperature, and has the general form:

$$U = U(T) = n \int c_\mathrm{v}(T)\,\mathrm{d}T + C \, , \tag{5}$$

where $C$ is an arbitrary constant. In Eq. (5), $n$ represents the number of moles of gas, and $c_V(T)$ is its molar specific heat at constant volume, which in principle can depend on temperature. Often, the molar specific heat of the ideal gas is assumed to be constant, and simply denoted as $c_V$. In this case, Eq. (5) is simplified to

$$U = U(T) = n\, c_V\, T + C. \tag{6}$$

In an ideal gas, pressure is related to temperature and volume by the well-known state equation:

$$p = n\, R\, T/V\,, \tag{7}$$

where $R$ is the universal gas constant. Equations (5) to (7) define the ideal gas. By convention, the gas pressure, $p$, is considered to be positive in compression. Accordingly, the work done on the gas to produce a change in volume d$V$ is given by

$$đW = -p\,\mathrm{d}V = -\,n\,RT\frac{\mathrm{d}V}{V}\ . \tag{8}$$

Finally, it is useful to recall the following four statements that are fundamental to classical thermodynamics.

(*i*) *The sum of thermal and nonthermal energy that a system can supply while keeping its internal energy constant is zero*.

This statement is an immediate consequence of the first law of thermodynamics in Eq. (1), which implies that

$$đW + đQ = 0 \qquad \text{if}\ \ \mathrm{d}U = 0. \tag{9}$$

In particular, this means that to absorb a net amount of energy, any material system must be capable of storing it as internal energy.

(*ii*) *Thermal equilibrium is the condition in which no net heat exchange occurs between two objects in thermal contact.*

(*iii*) *Any two systems at the same temperature are in thermal equilibrium with each other*.

This last statement, or an equivalent, is usually referred to as the **zeroth law of thermodynamics**.

Taken together, statements (*ii*) and (*iii*) mean that

(*iv*) *No heat exchange can occur between two systems in thermal contact if their temperatures are equal*.

It should be emphasized that systems in thermal equilibrium with each other need not be in absolute equilibrium; for example, two systems in thermal equilibrium at the same temperature can (and usually do) exchange mechanical energy when they interact with each other, if they are not in mechanical equilibrium. However, no heat exchange can take place between them during this process if their temperature remains unaltered.

In the following sections, we will examine the consequences of the above premises, proceeding in rigorous coherence with them.

## 3. Isothermal and Quasi-Isothermal Reversible Processes

To avoid ambiguities in the subsequent analysis, the following definition is introduced:

*A **strictly isothermal process** is a process in which* $\mathrm{d}T = 0$ *at any instant of the process.*

When there is no risk of confusion, a strictly isothermal process will simply be called an isothermal process.

From statement (*ii*) and the definition of a reversible process given in the previous section, it immediately follows that

$$đQ = 0 \tag{10}$$

in any reversible, strictly isothermal process. In other words, *all reversible, strictly isothermal processes are also adiabatic*.

From Equations (10) and (1), it can also be immediately concluded that the equation

$$\mathrm{d}U = đW \tag{11}$$

must apply to any system undergoing strictly isothermal reversible processes. That is, *the work absorbed or produced by any system in a strictly isothermal reversible process is equal to the change in the internal energy of the system.*

Strictly isothermal processes must be carefully distinguished from *quasi-isothermal processes*, which can be defined as follows:

*A **quasi-isothermal process** is a sequence of steps that, while not isothermal, are sufficiently small to maintain the temperature of the system close to a constant value within a small predefined tolerance.*

Since it contains non-isothermal parts, a quasi-isothermal process cannot be strictly isothermal. However, it may be reversible if all its constituent steps are performed reversibly. Regardless of whether it is reversible, the temperature fluctuations of the process will produce heat exchange between the system and its surroundings unless the system is insulated by adiabatic walls.

A widespread, though unfortunate, practice is to treat quasi-isothermal processes as strictly isothermal despite the fact that they produce very small or infinitesimal temperature variations around a specified temperature. However, these temperature changes signify that $\mathrm{d}T \neq 0$, meaning that a quasi-isothermal process is never isothermal ($\mathrm{d}T = 0$). A clear distinction between strictly isothermal and quasi-isothermal processes is fundamental to the theory of thermodynamics, although it may be less significant in some of its practical applications.

An important subclass of quasi-isothermal processes, referred to as pseudo-isothermal processes, will be discussed in Section 7.

**4. Isothermal Processes in Ideal Gases**

The isothermal condition, $\mathrm{d}T = 0$, imposes some stringent restrictions on the behavior of an ideal gas. The constitutive equation (5) for these gases implies that $\mathrm{d}U = 0$ if $T = \text{const}$. In this case, Eq. (11) is reduced to $đW = 0$, meaning that *no work can be done by/on an ideal gas in a reversible strictly isothermal process*.

By recalling Eq. (10), we can conclude that in the same processes,

$$đQ = đW = \mathrm{d}U = 0. \tag{12}$$

That is, every strictly isothermal reversible process for an ideal gas is, at the same time, adiabatic, workless, and occurs at constant internal energy.

The above conclusions follow from the fact that the internal energy of an ideal gas depends exclusively on temperature. This is a defining property of the ideal gas, and specifies that the gas can only store energy in thermal form. Any change in the internal energy of the gas corresponds to a change in its temperature, irrespective of the changes in volume that the gas may undergo during the process. Mechanical work can affect the internal energy of an ideal gas only by changing its temperature (i.e., through a non-isothermal process).

The situation is different for non-ideal gases. In this case, the gas particles are also subjected to forces due to long-range intermolecular interactions, and these forces depend on the distances between the particles and therefore on the volume of the gas. This causes the internal energy of the gas to depend not only on the temperature but also on the volume. In this scenario, a change in gas volume may lead to reversible isothermal ($\mathrm{d}T = 0$) processes in which $dU \neq 0$ and $đW \neq 0$ while still maintaining $đQ = 0$, as required for the isothermal process to be reversible; cf. Eq. (10).

## 5. Volume-Independence of the Entropy of an Ideal Gas

The entropy of a thermodynamic system is a state function whose differential, for reversible processes, is given by Eq. (3). The state of an ideal gas is fully determined by the variables $T$ and $V$. It is therefore expected that the entropy of an ideal gas should be a function of the form $S = S(T, V)$. This section analyzes the restrictions imposed on this function by the laws of thermodynamics, and demonstrates the existence of a rather strong restriction: the entropy of an ideal gas cannot depend on volume.

In support of this result, two independent proofs are presented. Proof 1 is based on the first and second laws of thermodynamics, combined with the third law. Proof 2 is a direct consequence of the properties of reversible and strictly isothermal processes, as introduced in the previous sections.

### Proof 1: Constant-volume reversible processes

Let us consider an ideal gas undergoing a reversible isochoric ($V$ = const.) process from an initial temperature $T_0$ to a final temperature $T$. Since the volume is constant, no work is done in the process. From the first law of thermodynamics, Eq. (1), we have

$$dU = đQ \tag{13}$$

at any instant of the process. From this equation and from Eq. (4), we can infer that in the same process,

$$\Delta S = \int_{T_0}^{T} \frac{dU}{T} \quad . \tag{14}$$

Since the internal energy of an ideal gas depends only on temperature, the right-hand side of Eq. (14) depends only on $T_0$ and $T$, and not on $V$. However, the same equation is valid for every value of $V$, $T_0$, and $T$. Hence, the difference $\Delta S = S(V, T) - S(V, T_0)$ must take the same value irrespective of the value of $V$. This implies that $S$ itself cannot depend on $V$ since, according to the third law of thermodynamics, it is independent of $V$ at a temperature of absolute zero. Thus, in an ideal gas, we have not only $U = U(T)$ but also $S = S(T)$.

This conclusion follows directly from the laws of thermodynamics, without reference to heat transfer in isothermal processes.

**Proof 2: Reversible strictly isothermal processes**

We begin by differentiating the functional relation $S = S(T, V)$, which expresses, in principle, the dependence of entropy on the variables describing the state of an ideal gas. This gives:

$$\mathrm{d}S = \frac{\partial S}{\partial T}\,\mathrm{d}T + \frac{\partial S}{\partial V}\,\mathrm{d}V \, . \tag{15}$$

By substituting this into Eq. (3), we find that the following relation

$$đQ = T\left(\frac{\partial S}{\partial T}\,\mathrm{d}T + \frac{\partial S}{\partial V}\,\mathrm{d}V\right) \tag{16}$$

must be satisfied in any reversible process of an ideal gas. In particular, Eq. (16) must be met in isothermal reversible processes. However, in this case, we have $\mathrm{d}T = 0$ because the process is isothermal. Moreover, $đQ = 0$ due to the fact that the process is reversible, and must therefore occur in thermal equilibrium with the surroundings (cf. Eq. (10)). However, the quantity $\mathrm{d}V$ may be non-zero depending on the particular isothermal process that is considered. Hence, for Eq. (16) to hold in any reversible isothermal process, we must have

$$\frac{\partial S}{\partial V} = 0 \, . \tag{17}$$

Since it is valid for all values of $T$ and $V$, this equation requires that the entropy of an ideal gas cannot depend on volume. That is $S = S(T)$.

The explicit form of the entropy for an ideal gas is derived in the next section.

## 6. Explicit Expression for the Entropy of the Ideal Gas

To determine the entropy function of an ideal gas, it is sufficient to consider constant-volume heating and cooling processes, since, as shown above, the entropy of this gas depends only on temperature. These processes will be referred to here as *simple heating* and *cooling*. They can be carried out reversibly if they are conducted sufficiently slowly to ensure that the gas remains isothermal throughout its volume.

Let us then consider a reversible simple heating process from an initial temperature $T_0$ to a final temperature $T$. In any infinitesimal time interval, $\mathrm{d}t$, in the process, we have $\mathrm{d}W = 0$ because $\mathrm{d}V = 0$. Eq. (13) therefore applies to this process. From Eq. (5), we know that

$$\mathrm{d}U = n\, c_V(T)\, \mathrm{d}T. \tag{18}$$

From this and from Eqs. (3) and (13), we obtain

$$\mathrm{d}S = \frac{\mathrm{d}U}{T} = \frac{n\, c_\mathrm{V}(T)}{T}\mathrm{d}T \,. \tag{19}$$

Integrating this equation from $T_0$ to $T$ gives

$$S(T) - S(T_0) = n \int_{T_0}^{T} \frac{c_\mathrm{V}(T')}{T'}\mathrm{d}T' \;, \tag{20}$$

where $T'$ denotes the integration variable in the interval $[T_0, T]$. Equation (20) gives the expression for the entropy of an ideal gas in its complete generality. For the actual calculation, the dependence of $c_\mathrm{V}$ on temperature must be made explicit.

At temperatures far above absolute zero and sufficiently low pressures, the molar heat capacity of the ideal gas can be assumed to be constant and simply denoted as $c_\mathrm{V}$. In this case, Eq. (20) is simplified to

$$S(T) = n\, c_\mathrm{V} \ln T + const. \tag{21}$$

The additive constant that appears in this equation may be fixed by specifying the value of entropy for a particular value of $T$. In most cases, however, this constant can safely be ignored.

Equation (21) and the more general Eq. (20) are expressions for the entropy of an ideal gas that follow from a rigorous application of the principles of classical thermodynamics. They show that the entropy of an ideal gas does not depend on volume, unlike in the expression

$$S = S(T,V) = n\, c_\mathrm{V} \ln T + n\, R \ln V + const. \tag{22}$$

that is generally used in the literature. As demonstrated in the previous section, the presence of the volume-dependent term is incompatible with the restrictions imposed by a strict adherence to the principles of classical thermodynamics.

## 7. Validity of the Standard Isothermal Formulae: Pseudo-Isothermal Processes

The conclusion that a strictly isothermal reversible process of an ideal gas cannot involve any exchange of heat or work (Section 4) presents a formal paradox when contrasted with standard textbook expressions. This apparent contradiction is resolved in this section by recognizing that traditional isothermal formulae do not strictly apply to strictly isothermal processes but rather represent an operational limit that the gas can attain through processes that, while not strictly isothermal, keep the gas temperature as close to a constant value as desired. These are the pseudo-isothermal processes discussed below.

### A. Definition of a reversible pseudo-isothermal process

A reversible pseudo-isothermal process at a nominal temperature $\bar{T}$ consists of an alternating sequence of reversible process or steps of the following two types:

1. *Reversible adiabatic steps*, in which the gas performs/receives work and undergoes temperature variations;
2. *Reversible constant-volume heating or cooling steps*, in which the gas does no work but exchanges heat with its surroundings to reach a prescribed temperature.

To qualify as pseudo-isothermal, the process must satisfy the following additional conditions:

3. A pseudo-isothermal process should start and end with a constant-volume heating or cooling step starting or ending at temperature $\bar{T}$.
4. All adiabatic steps must extend above and below the temperature $\bar{T}$ so that the weighted average temperature of the step, with $1/V$ as the weight function, coincides with $\bar{T}$.

More specifically, let $(T_i, V_i)$ and $(T_{i+1}, V_{i+1})$ be the values of temperature and volume at the beginning, $i$, and end, $i+1$, of the generic adiabatic step. Condition 4 demands that these values be so chosen as to satisfy the relation

$$\int_{V_i}^{V_{i+1}} T \frac{1}{V} \mathrm{d}V = \bar{T} \int_{V_i}^{V_{i+1}} \frac{1}{V} \mathrm{d}V \tag{23}$$

at each adiabatic step of a pseudo-isothermal process. It is not precluded, of course, that $(T_i, V_i)$ and $(T_{i+1}, V_{i+1})$ may vary from step to step (see, for example, Fig. 1). The positivity of the variable $V$ guarantees that the above equation is free from singularities.

Incidentally, if the adiabatic step $[V_i, V_{i+1}]$ is sufficiently small compared to $V$, the weighted average temperature mentioned in condition 4 can be approximated by the mean value of $T$. Moreover, the graph of $T$ vs. $V$ for the specified step approaches a straight-line segment that intersects the isotherm $T=\bar{T}$ at the midpoint between $V_\mathrm{i}$ and $V_{\mathrm{i}+1}$.

None of the steps of a pseudo-isothermal process are isothermal. However, if the steps are sufficiently small and correctly sequenced, the gas temperature can be kept arbitrarily close to the constant value $\bar{T}$ during the entire process (see Fig. 1, inset). In this way, a reversible pseudo-isothermal process simulates a strictly isothermal reversible process at temperature $\bar{T}$, while allowing the exchange of work and heat between the system and its surroundings.

A pseudo-isothermal process expands or compresses the gas, depending on whether it proceeds in the direction of increasing or decreasing $V$. The adiabatic steps of a pseudo-isothermal process determine the work that is done. The constant-volume steps control the temperature of the gas, but do not involve any work.

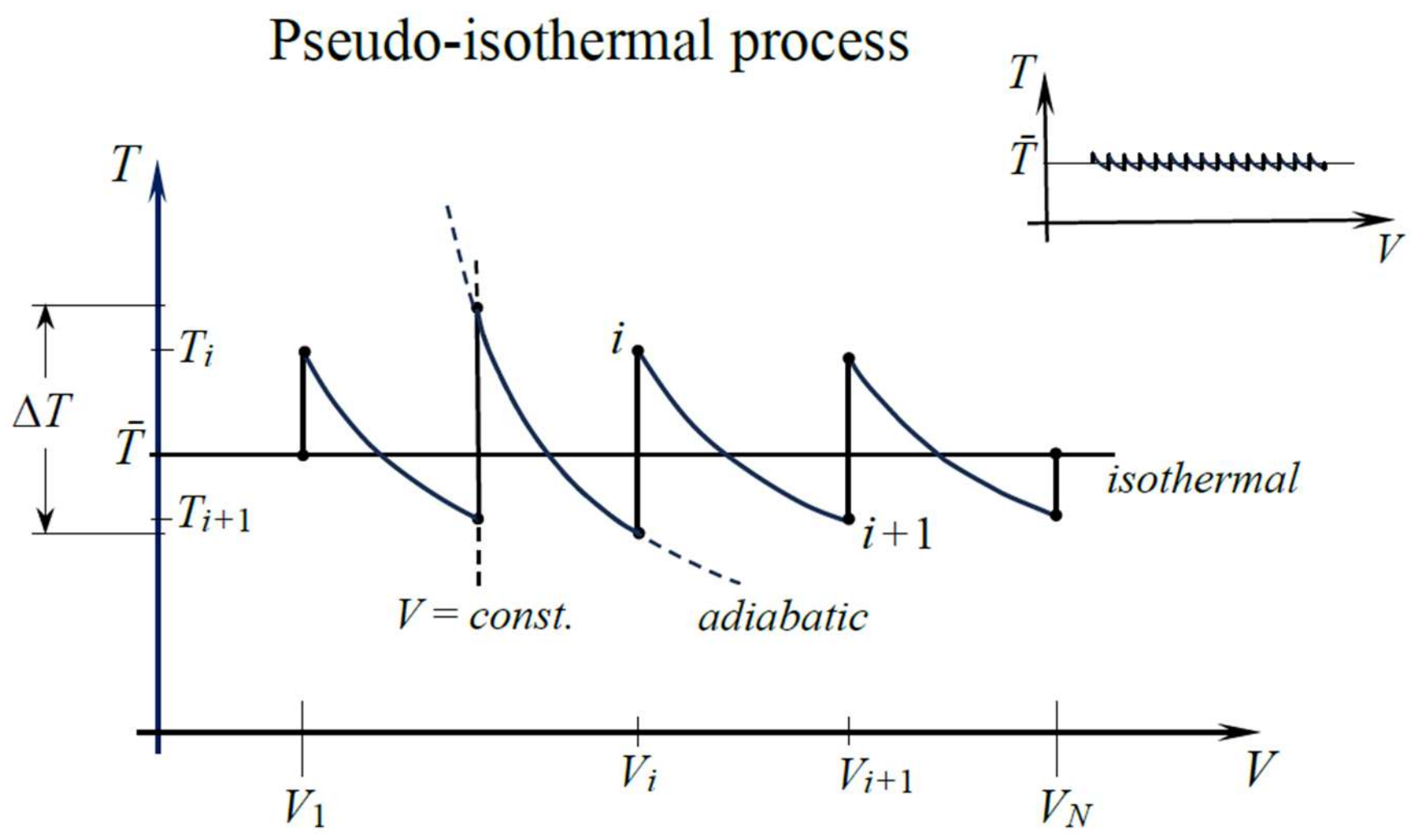

**Figure 1**. Graphical representation of a pseudo-isothermal process resulting from a sequence of work-producing adiabatic steps alternating with workless steps of heat exchange at a constant volume. By controlling the temperature variation at each step, the gas temperature can be kept arbitrarily close to the nominal constant value $\bar{T}$ over the whole process *(inset)*.

### B. Work performed by an ideal gas in a reversible pseudo-isothermal process

Consider the reversible pseudo-isothermal expansion of an ideal gas from an initial volume $V_1$ to a final volume $V_N$, with a nominal constant temperature $\bar{T}$ (Fig. 1). In each adiabatic step of this process, say from point $i$ to point $i$+1, the work performed by the gas is

$$W_i = -\int_{V_i}^{V_{i+1}} p\,\mathrm{d}V \,. \tag{24}$$

Using Eq. (7) and applying Eq. (23), this equation can be written as

$$W_i = -\int_{V_i}^{V_{i+1}} \frac{nRT}{V}\,\mathrm{d}V = -\,n\,R\,\bar{T} \int_{V_i}^{V_{i+1}} \frac{1}{V}\,\mathrm{d}V = \;n\,R\,\bar{T}\,\ln\frac{V_i}{V_{i+1}} \;, \tag{25}$$

By summing the contributions $W_i$ from each adiabatic step, Eq. (25), the work done on the ideal gas in a pseudo-isothermal process can be easily calculated as:

$$W = \sum_{i=1}^{N-1} W_i = n\,RT \sum_{i=1}^{N-1} \ln \frac{V_i}{V_{i+1}} \ . \tag{26}$$

Since the sum of logarithms is equal to the logarithm of the product of their arguments, Eq. (26) can be simplified to

$$W = nRT \ \ln \frac{V_1}{V_N} = -nRT \ \ln \frac{V_N}{V_1} \,. \tag{27}$$

This is exactly the classical formula for the work done by an ideal gas in a reversible isothermal expansion or compression process (see, for example, p. 23 of Ref. [9], or p. 57 of Ref. [10]).

In summary, no work can be done by an ideal gas in a reversible strictly isothermal process (Section 4). However, the work done by the ideal gas in a reversible pseudo-isothermal process (where, although not strictly isothermal, the temperature of the gas is kept as close as desired to any given constant temperature) coincides with the work that is traditionally (and less rigorously) assigned to isothermal reversible processes (expansions or compressions).

### C. Pseudo-isothermal heat transfer

In a pseudo-isothermal process, the initial and final temperatures of the process can always be made to coincide with the nominal value by inserting a constant-volume heating/cooling step starting/ending at that temperature. This is the case in the example shown in Figure 1, and will be assumed to hold in all of the analyses below.

For an ideal gas, this choice means that the pseudo-isothermal process will produce no final change in the internal energy of the gas, as this energy depends only on temperature (see Eqs. (5) and (6)). As a consequence, integrating Eq. (1) over the complete pseudo-isothermal process gives

$$Q = -\,W \ . \tag{28}$$

In other words, the total amount of heat absorbed in the process is equal to the work performed by the gas. Of course, a similar result is not true at every instant of the process; in fact, in a reversible pseudo-isothermal process, we always have $đQ \neq 0$ and $đW = 0$ in the constant-volume steps, and $đQ = 0$ and $đW \neq 0$ in the adiabatic ones.

Equations (27) and (28) show that in an ideal gas, a reversible pseudo-isothermal process results in the same mathematical expressions for work and heat transfer as those traditionally associated with isothermal processes. This means that substituting isothermal processes with pseudo-isothermal processes does not affect the fundamental structure of classical

thermodynamics—including the Carnot cycle and the theorems on the efficiency of thermal engines. The rigorous notion of pseudo-isothermal processes, therefore, helps us eliminate the conceptual need to assume that heat can flow between a system and its environment in isothermal equilibrium conditions.

Finally, it must be observed that in an ideal gas, a reversible pseudo-isothermal process is not strictly isentropic. Temperature fluctuations are accompanied by entropy fluctuations, as is evident from Eqs. (20) or (21). This fact prevents the process from being strictly isentropic. However, since the process does not lead to any final change in the temperature of the gas, it does not result in any final change in its entropy (see Eqs. (20) and (21)).

**8. Free Expansion of Ideal Gases**

In a free expansion, a gas expands into a vacuum without performing external work. This means that at each instant of the process, $đW = 0$ and, consequently,

$$\mathrm{d}U = đQ\,, \tag{29}$$

as follows from Eq. (1). If the free expansion is executed adiabatically, Eq. (29) yields $\mathrm{d}U = 0$. For an ideal gas, this means that the free expansion must be isothermal, as follows from Eq. (5).

Conversely, the isothermal free expansion of an ideal gas must be adiabatic, since in this gas the condition $T =$ const. implies $\mathrm{d}U = 0$ and hence $đQ = 0$ from Eq. (29). The adiabaticity of the process (i.e., the condition that $đQ = 0$ at all times of the process) is therefore both necessary and sufficient for the free expansion of an ideal gas to be isothermal.

As shown in Section 5, the entropy of an ideal gas is uniquely determined by its temperature. Since the temperature of the gas remains constant during an adiabatic free expansion, it must be concluded that in the same process,

$$\mathrm{d}S = 0\,. \tag{30}$$

That is, the free expansion of the ideal gas is also isentropic. Furthermore, because the process is adiabatic, we know that $đQ = 0$. From Eq. (3), we can therefore conclude that *the free expansion of an ideal gas is, in fact, a reversible process.*

While this conclusion directly challenges the deeply entrenched textbook paradigm that classifies all free expansions as inherently irreversible, this divergence is actually an artifact of importing volume-dependent definitions of statistical entropy into a purely macroscopic framework.

To show how the expanded gas can actually be reversibly returned to its pre-expansion state, we first note that thermodynamic reversibility refers to the possibility of restoring the gas to its initial state without leaving any change in the environment. Consider then an ideal gas undergoing a quasi-static free expansion in a vacuum from volume $V_I$ to volume $V_F > V_I$. Let $T$ be the gas temperature in the process. To analyze the process in classical thermodynamic terms, the gas expansion must be treated as a sequence of equilibrium states, with the variables $T$ and $V$ completely determining the state of the gas at any instant of the process.

A quasi-static free expansion such as this can be produced in practice: it is sufficient to place the gas into a chamber of volume $V_I$, adjacent to a sequence of small-volume vacuum chambers and separated from them by removable, weightless, and frictionless partitions. The partitions are then sequentially removed until the volume $V_F$ is reached. By adjusting the removal times and reducing the size of the small-volume chambers, the resulting quasi-static process can approximate any free expansion of the gas with the desired degree of accuracy.

Once the expanded state ($T$, $V_F$) is reached, the gas can be returned to its initial state ($T$, $V_I$) via the following reversible three-step sequence:

1. **Isochoric cooling**. By keeping its volume constant at the final value $V_F$, the expanded gas is reversibly cooled from the temperature $T$ to a baseline temperature $T_0$ approaching absolute zero. No work is done in the process, but the gas releases a certain amount of heat.

2. **Isothermal compression at low temperature.** At the temperature $T_0$, the ideal gas is compressed isothermally and reversibly to restore its volume to the initial value $V_I$. Being isothermal, the process maintains the internal energy of the gas unaltered ($dU = 0$). Moreover, by choosing the value of $T_0$ sufficiently close to absolute zero, both the mechanical work expended and the heat produced during this step can be made arbitrarily small, as dictated by the vanishing pressure of the gas at absolute zero and the first law of thermodynamics, Eq. (1).

3. **Isochoric heating.** The gas is reversibly heated at constant volume $V_I$ from $T_0$ back to its original temperature $T$. In this step, the gas performs no work and absorbs the exact amount of heat that was released in the initial cooling step. This step returns the gas to its initial state ($T$, $V_I$).

The cooling–compression–heating sequence described above reverses the adiabatic free expansion of the ideal gas. The work and heat exchanged with the external environment during the low-temperature compression step can be rendered smaller than any predefined value as $T_0$ tends to 0 K. The considered sequence confirms, therefore, that the adiabatic free expansion of an ideal gas is a logically reversible process within the strict boundaries of macroscopic classical thermodynamics.

### 9. Ideal Gas Free Energy and Relevance of Maxwell's Relations

For simplicity, in this section we will refer to the case of ideal gases with constant specific heat. Similar conclusions apply to temperature-dependent specific heat. The Helmholtz free energy, $\Psi$, for an ideal gas with constant specific heat is easily obtained from Eqs. (6) and (21):

$$\Psi = \Psi(T) := U - T\,S = n\,c_V\,(T - T\ln T) + C. \tag{31}$$

Accordingly,

$$\mathrm{d}\Psi = n\,c_V\left[1 - \left(\ln T + T\frac{1}{T}\right)\right]\mathrm{d}T = -\,n\,c_V \ln T\,\mathrm{d}T\,. \tag{32}$$

For isothermal processes, $\mathrm{d}T = 0$. Hence, from Eq. (32) we infer that

$$\mathrm{d}\Psi\big|_{T=\mathrm{const}} = 0\,. \tag{33}$$

For the same processes, the definition given by Eq. $(31)_2$ and the first and second laws of thermodynamics, Eqs. (1) and (2), yield

$$\mathrm{d}\Psi\big|_{T=\mathrm{const}} \geq \mathrm{d}W\ , \tag{34}$$

where the equality sign applies if the process is reversible. By virtue of Eq. (33), this relation implies that for any isothermal process of an ideal gas, the relation

$$\mathrm{d}W \leq 0 \tag{35}$$

must apply, the equality sign being valid when the process is reversible.

Relation (35) allows us to conclude that under isothermal conditions, an ideal gas cannot absorb work from the surroundings ($\mathrm{d}W > 0$), regardless of whether or not the process is reversible. The same relation also indicates that $\mathrm{d}W = 0$ in all reversible isothermal processes of the same gas. These conclusions are consistent with the previous finding that in a strictly isothermal reversible process, the ideal gas can neither absorb nor supply work (Section 4). Consequently, free expansions are the only isothermal processes that an ideal gas can perform reversibly (Section 8).

As is evident from relation (35), an ideal gas can supply work isothermally ($\mathrm{d}W < 0$), but only in an irreversible way. The work supplied must at any instant be equal to the heat that the gas absorbs from the surrounding environment; this follows from Eq. (1), since under isothermal conditions, $\mathrm{d}U = 0$ for ideal gases. The required heat must come from a source at a higher temperature than the gas, since no heat transfer is possible in thermal equilibrium conditions. This makes the process are thermodynamically irreversible.

Let's now refer to the applicability of Maxwell's relations to ideal gases. For any reversible processes, we have $đQ = T\,đS$. In this case, the first law in Eq. (1) yields:

$$\mathrm{d}U = T\,\mathrm{d}S - p\mathrm{d}V\ . \tag{36}$$

Using Eq. $(31)_2$, this equation can be written as

$$\mathrm{d}\Psi = -\,S\,\mathrm{d}T - p\,\mathrm{d}V, \tag{37}$$

which expresses the same balance (36) in terms of Ψ rather than *U*. Equation (37) is valid regardless of the form of the functions involved.

In particular, for systems in which $\Psi = \Psi(T, V)$, we have

$$\mathrm{d}\Psi = \frac{\partial \Psi}{\partial T}\mathrm{d}T + \frac{\partial \Psi}{\partial V}\mathrm{d}V\ . \tag{38}$$

If we insert this equation into Eq. (37), we can immediately conclude that

$$S = -\frac{\partial \Psi}{\partial T} \tag{39}$$

and

$$p = -\frac{\partial \Psi}{\partial V}. \tag{40}$$

By taking the partial derivatives of Eqs. (39) and (40) with respect to *V* and *T*, respectively, and by exploiting the symmetry property of the second-order mixed partial derivatives of Ψ, it can be concluded that

$$\left(\frac{\partial S}{\partial V}\right)_T = \left(\frac{\partial p}{\partial T}\right)_V, \tag{41}$$

which is the well-known Maxwell's relation based on the Helmholtz free energy of the system.

However, it is apparent from Eq. (31) that the free energy of an ideal gas depends solely on *T*. Hence, for an ideal gas, Eq. (38) is reduced to

$$\mathrm{d}\Psi = \frac{\mathrm{d}\Psi}{\mathrm{d}T}\mathrm{d}T\ . \tag{42}$$

In this case, the functions *S* and *U*, and thus Ψ, do not depend on volume, which makes their partial derivatives with respect to *V* meaningless. As a consequence, any conclusions drawn from the symmetry of the mixed second-order partial derivatives of the functions Ψ, *S*, and *U*, including Maxwell's equation (41), are not applicable to the ideal gas.

Similarly, Eq. (40) cannot be applied to an ideal gas, because in this case $\Psi = \Psi(T)$. This does not pose a problem, because Eq. (7) is a constitutive hypothesis that is independent of any assumption regarding $U$, $S$, and $\Psi$. It simply states that at a given temperature, the equilibrium value of $p$ is proportional to the number of gas particles per unit volume. Conversely, Eq. (37) is applicable to ideal gasses as it only represents an alternate formulation of the general energy balance found in Eq. (1) for reversible process of the same gasses.

**10. Entropy of an Ideal Gas in Kinetic Theory and Statistical Mechanics**

Maxwell's kinetic theory of gases offers a statistical mechanical interpretation of several properties of an ideal gas, including temperature, pressure, specific heat and internal energy. In Maxwell's theory, the temperature of an ideal gas is related to the average kinetic energy, $\overline{K}$, of the gas particles by the relation

$$\overline{K} = \frac{1}{2} m \, \overline{v^2} = \frac{3}{2} k \, T \ . \tag{43}$$

Here, $m$ denotes the mass of a gas particle, $v$ is the particle velocity, and $\overline{v^2}$ is the average value of $v^2$, taken over all the gas particles. The constant $k = R/N_A$ is Boltzmann's constant, where $R$ and $N_A$ are the universal gas constant and the Avogadro number, respectively. The temperature of the gas is therefore given by

$$T = \frac{m}{3k} \overline{v^2} \ . \tag{44}$$

In the same theory, the internal energy of an ideal gas coincides with the total kinetic energy of its particles, and can therefore be expressed as

$$U = N \, \overline{K} = \frac{1}{2} N \, m \, \overline{v^2} = \frac{3}{2} N \, k \, T = \frac{3}{2} n \, R \, T \ , \tag{45}$$

where $N = n \, N_A$ indicates the total number of the gas particles (cf., for example, p.132 of Ref. [10], or p. 267 of Ref. [11]). The molar specific heat of the ideal gas is then given by:

$$c_V = \frac{3}{2} R \tag{46}$$

(cf., for example, p. 267 of Ref. [12]).

In its classical formulation, Maxwell's kinetic theory does not consider entropy. However, from the above equations, expressing the entropy of the gas as a function of $\overline{v^2}$ is rather straightforward. We need only observe that in a reversible simple heating or cooling process, inserting Eq. $(45)_2$ into Eq. (17) yields:

$$đQ = \frac{1}{2} N \, m \, \mathrm{d}\left(\overline{v^2}\right) \ . \tag{47}$$

When this equation and Eq. (44) are introduced into Eq. (3), we obtain, after minor rearrangements,

$$\mathrm{d}S = \frac{3}{2} N\, k \,\frac{1}{\overline{v^2}}\, \mathrm{d}\left(\overline{v^2}\right) = \frac{3}{2}\, n\, R \,\frac{1}{\overline{v^2}}\, \mathrm{d}\left(\overline{v^2}\right). \tag{48}$$

Integrating this with respect to $\overline{v^2}$ produces

$$S = \frac{3}{2}\, n\, R \ln\left(\overline{v^2}\right) + \mathrm{C}, \tag{49}$$

where the quantity C is an integration constant. Equation (49) gives the expression for the entropy of an ideal gas that is implied by the kinetic theory.

Through a few elementary substitutions, this expression can also be written as

$$S = \frac{3}{2}\, n\, R \ln\left(\frac{m}{3k}\overline{v^2}\right) - \frac{3}{2}\, n\, R \ln\left(\frac{m}{3k}\right) + \mathrm{C} = \frac{3}{2}\, n\, R \ln\left(\frac{m}{3k}\overline{v^2}\right) + \mathrm{C}', \tag{50}$$

where C′ is another constant, given by

$$\mathrm{C}' = \mathrm{C} - \frac{3}{2}\, n\, R \ln\left(\frac{m}{3k}\right). \tag{51}$$

Taken together, Eqs. $(50)_2$, (44), and (46) indicate that the entropy of ideal gases, as implied by the kinetic theory of gases, is consistent with the expression in Eq. (21) that was derived in Section 6.

Apparently, Eqs. (49) and (50) have not received the consideration they deserve. Maxwell's kinetic theory stimulated an intense search for a mechanical interpretation of thermodynamics, which ultimately led to modern statistical mechanics. The latter theory, however, was based on the assumption that entropy was given by the Planck–Boltzmann equation:

$$S = S(N,U,V) = k \ln \Omega\,, \tag{52}$$

where $\Omega = \Omega(N,U,V)$ is the number of microstates compatible with the specified macrostate of the gas particles, and is determined by the number ($N$) of particles, the total energy ($U$), and the total volume ($V$) occupied by the gas.

In Eq. (52), the entropy of the ideal gas depends on its volume, despite the fact that the internal energy of the gas, and thus its thermal energy content, are volume-independent. The dependence on $V$ makes Eq. (52) inconsistent with the thermodynamic entropy of the ideal gas, Equations (20) or (21), and its kinetic theory counterpart, Eq. (49). This indicates that the entropy of statistical mechanics, given by Eq. (52), is actually a different quantity from the entropy that is introduced in classical thermodynamics, see also References [6] and [7].

The same conclusion applies to all generalizations of Eq. (52) in which the dependence of volume is due to statistics rather than thermodynamics. This is the case, for instance, for Gibb's formula:

$$S = -k \sum_i p_i \ln p_i \,, \tag{53}$$

where $p_i$ represents the probability of finding the system in the $i$-th microstate. The variable $p_i$ is volume-dependent because the number of positions available to the particles depends on volume. Equation (53) is used to introduce entropy in the fields of quantum mechanics and information theory.

## 11. Conclusions

A rigorous axiomatic treatment of macroscopic thermodynamics shows that the thermodynamic entropy of an ideal gas is independent of volume, in accordance with the fact that its internal energy is a function of temperature alone. This finding does not disprove the practical validity of classical thermodynamics but clarifies its formal architecture. By distinguishing between strictly isothermal and *pseudo-isothermal* processes, this work provides a rigorous explanation for why standard isothermal formulae are exact in applied contexts, such as the analysis of thermal engines and the Carnot cycle, where processes are operationally pseudo-isothermal.

Furthermore, the independence of the macroscopic entropy of an ideal gas from the volume fits seamlessly with the microscopic standpoint provided by Maxwell's kinetic theory of gases. The apparent discrepancy between this result and the volume-dependent Planck–Boltzmann formula highlights a fundamental conceptual difference: thermodynamic entropy and statistical microstate-counting represent different physical properties of a system. The statistical dependence on volume, arising from spatial distributions of particles, concerns the description of microscopic configurations rather than the strict macroscopic constraints of classical thermal equilibrium.

Recognition of this difference resolves long-standing formal ambiguities at the boundary between macroscopic thermodynamics and statistical mechanics. It also provides classical thermodynamics with an independent, paradox-free axiomatic basis, liberating its fundamental concepts from the structural limitations deriving from the models of microscopic disorder and re-establishing it as a self-contained, mathematically rigorous macroscopic science.